\documentstyle[12pt]{article}
\input psfig
\long\def\comment#1{}
\begin{document}
\title{The Cubic Public-Key Transformation}
\author{Subhash Kak}
\date{}
\maketitle

\begin{abstract}
We propose the use of the cubic transformation for public-key applications
and digital signatures. Transformations modulo a prime $p$ or a 
composite $n=pq$, where
$p$ and $q$ are primes,
are used in such a fashion that each transformed
value has only 3 roots that makes it a more efficient transformation than 
the squaring
transformation of Rabin, which has 4 roots.
Such a transformation, together with additional tag information, makes 
it possible
to uniquely invert each transformed value.
The method may be used for other exponents as well.

\end{abstract}

\thispagestyle{empty}

\section{Introduction}
Certain many-to-one mappings, when used together with additional 
side information, may 
be uniquely inverted and used
for  public-key cryptography and digital signatures. 
Given
message $m$ in $Z_n$,  we propose the use of
the encryption function $c = m^3 \bmod n$ under suitable restrictions
on $m$ and $n$. We will be primarily interested in the case
where 3 divides the Euler totient function
 $\phi (n)$, but 9 does not.

For secure communication, Alice transmits to Bob $c$ together with 
side information which may be sent in an encrypted form or 
published.

This transformation is more efficient than the 
mapping 
$c = m^2 \bmod n $ proposed by Michael Rabin [1] that
has the virtue of being provably at most as intractable as the factorization
of $n$. 
In his article, Rabin did consider generalization beyond
the squaring function, but 
he dismissed
the use of higher powers for 
their relative inefficiency. Specifically,
he spoke of the 9
cube roots in the cubic mapping, having
implicitly taken $n$ to be
divisible by $9$,
which made it less
efficient than the squaring transformation with its four roots.

The $n$ that we choose ensures only three cube roots for each value making 
it more
efficient than the squaring transformation.
We also consider extensions of our side
information scheme to the case when $n$ is divisible by $9$.
Given its easy implementation, the proposed cubic transformation lends 
itself to
public key applications.
Specifically,

\begin{itemize}
\item It makes additional exponents available in exponentiation transformations [2,3]
and for key-distribution systems that use powers of primitive roots.
\item It makes it possible to use a small exponent to send information and
implement digital signatures.
\item It allows the generation of probability events based on whether the
parity of the bit chosen by the
recipient is the same as the one picked at the sending point.
\end{itemize}

\section{The properties of the cubic transformation}

Consider first the transformation $c = m^3 \bmod p$,
where $p= 3k + 1$ is prime, and $p=3 \bmod 4$. 
Since $p-1$ and $3$ have at least one common divisor
(namely 3),
this transformation is not one-to-one, and because of
the cubing operation
3 different values of $m$ map to the same $c$.

If $p-1$ is not divisible by $9$, one can obtain one of the three values 
of $c^{1/3}$ by means of an inverse exponentiation operation:

\begin{equation}
c^{1/3} = \left\{ \begin{array}{ll}
 	c^{\frac{p+2}{9}} & \mbox{if $p-1  \bmod 9 =6$ }\\
 	c^{\frac{2p+1}{9}} & \mbox{if $p-1 \bmod 9 = 3$}
 		\end{array}
		\right. 
\end{equation}

The proof of this assertion is in the fact that $c^{p-1} = 1$ by Fermat's
Little Theorem and
$c^{a(p-1)+3} = c^3$. 
Consequently,
$c^{\frac{a(p-1)+3}{9}} = c^{1/3}$. 
If $p-1 \bmod 9  = 6$, then ${p+2}$ is divisible by $9$;
if $p-1 \bmod 9 = 3 $,
then $2(p-1) +3 $ is divisible by $9$.

An inverse cannot be obtained
by exponentiation if $p-1$ is divisible by $9$.

The cube roots of $1$ are $1$, $\alpha$ and $\alpha^2$ 
because if $\alpha$ is a root, so is its square. 
This implies that if one of the cube roots of $c$ is $m$, the
others 
must be $ m \alpha$ and $ m \alpha^2$.

The three cube roots of 1 may be obtained by solving the equation:

\begin{equation}
\alpha^3 - 1 = 0
\end{equation}

Apart from the obvious solution of $1$, the other two solutions 
are obtained by solving $\alpha^2 + \alpha + 1 = 0$, which is
possible if the square root
of $\surd \overline{p-3}$ exists. 
This may be determined by the use of Euler's criterion that $b$ is square
modulo $p$ if and only if
\[ b^{(p-1)/2} = 1 \bmod p \]

Since $p = 3  \bmod 4$, 
the square root $a^{1/2} = {a}^{\frac{p+1}{4}}$, and we can write,

\begin{equation}
\alpha = \frac{-1 \pm \surd \overline{1-4}}{2}  
= \frac{1}{2} (-1 \pm (p-3)^{\frac{p+1}{4}})  
\end{equation}

\noindent
Since the two roots can also be written as $\alpha$ and $\alpha^2$, the 
value of $\alpha$ may be equivalently expressed as:

\begin{equation}
\alpha = \frac{-1 + (p-3)^{\frac{p+1}{4}}}{-1 - (p-3)^{\frac{p+1}{4}}  }
\end{equation}

\noindent
If the square root of $(p-3)$ does not exist, one must use a probabilistic
method to determine $\alpha$.

Note that $1+\alpha + \alpha^2 = 0 \bmod p$. This follows from the fact that
$(1+\alpha + \alpha^2)^3 = 1$. This means that the value of $\alpha^2$ may be 
obtained by subtraction.

The complexity of this computation is no more than that of exponentiation.

\vspace{0.1in}
\noindent
{\bf Example 1.} Consider $c = m^3  \bmod 31$.

To compute $\alpha$, we must
first find $\surd \overline{-3} = \surd \overline{28}$.
This can be reduced by using the transformation $28^{\frac{31+1}{4}} =
28^8$ which is equal to $20$.
Therefore, $\alpha = -21/2,~ 19/2$, which
may be reduced to $5$ and $25$. 
The cube roots of $1$ are, therefore, 1, 5, 25.
We may pick one of these, say 5, as  $\alpha$.

Let Alice choose $m=7$. She computes $m, ~m\alpha, ~m\alpha^2$, which are
4, 20, 7. Rearranging them in order, she gets 4, 7, 20 that correspond
to the ranks 1, 2, 3, and thus the
side information related to her choice of $m$ is 2.

Alice sends $c= 7^3 \bmod 31 = 2$, together with the rank information
of 2 (that is either published or sent in an encrypted form separately) to Bob. 
Using equation (1), Bob finds that
one of its cube 
roots is $2^7 ~\bmod ~31= 4$. The other
cube roots will be $4 \times 5 = 20 ~\bmod ~31$, and $20 \times 5 = 7  ~\bmod ~31$.
The side information then helps him pick $m=7$ as the message sent by
Alice.

\vspace{0.1in}
\noindent
{\bf Table 1}: Mapping for p = 31, $\alpha =5$
\vspace{0.1in}

\begin{tabular}{||c|c||} \hline
 Values of $m$, $m\alpha$, $m\alpha^2$ & The output $c$ \\ \hline
1, 5, 25 & 1 \\ \hline
2, 10, 19 & 8 \\ \hline
3, 15, 13 & 27 \\ \hline
4, 20, 7 & 2 \\ \hline
6, 30, 26 & 30 \\ \hline
8, 9, 14 & 16 \\ \hline
11, 24, 27 & 29 \\ \hline
12, 29, 21 & 23 \\ \hline
16, 18, 28 & 4 \\ \hline
17, 23, 22 & 15 \\ \hline
\end{tabular}

\section{Cubic transformation modulo a composite number}

We now consider the cubic transformation modulo $n = pq$, where $\phi (n) = (p-1) (q-1)$ is divisible by $3$
but not $9$. Then the inverse is given by:

\begin{equation}
c^{1/3} = \left\{ \begin{array}{ll}
 	c^{\frac{\phi (n)+3}{9}} & \mbox{if $\phi (n) \bmod 9 = 6$}\\
 	c^{\frac{2\phi (n)+3}{9}} & \mbox{if $\phi (n) \bmod 9 = 3$}
 		\end{array}
		\right. 
\end{equation}

The method of inverse transformation is identical to that in the
previous section, excepting that under certain circumstances the computation
of $\alpha$ will require solving equation (2) modulo the
two prime factors of $n$ and then combining them to get
the answer modulo $n$.
Both Alice and Bob know the factors of $n$, but the eavesdropper does not.

\vspace{0.1in}
\noindent
{\bf Example 2.} Consider $c = m^3 \bmod 77;~ p=7,~q=11$. 
To find the value of $\alpha$ by the method of equation (3), we must
obtain $\surd\overline{p-3} = \surd \overline {74}$. 
But 74 is not square modulo 77, therefore, 
we find the solution
directly by the Chinese Remainder Theorem (CRT), by solving:

$\alpha^3 -1 = 0$

\vspace{0.1in}
\noindent
separately for the two moduli 7 and 11. We can try to
obtain these solutions 
using the method of equation (3).
Note that only one new solution is required to fix $\alpha$ and, therefore,
it is not essential that solutions to both the primes are found.

For p=7, $p-3 = 4 $ is square modulo 7, and, therefore,
the solutions are easily obtained: 

\[\alpha_p = 2, 4\]

But 8 is not square modulo 11, therefore, equation (3) cannot be used for
it. We, therefore, merely use the obvious solution:

\[\alpha_q =1 \]

Combining, using the CRT,

\begin{equation}
\alpha = \alpha_p \times q \times ||q^{-1}||_{p} + \alpha_q \times p \times ||p^{-1}||_{q} \bmod pq
\end{equation}

\noindent
where $||x||_y = x \bmod y$.

Since $7^{-1} \bmod 11 = 8$, and $11^{-1} \bmod 7 = 2$, we have:

\[ \alpha_1 = 2 \times 11 \times 2 + 1 \times 7 \times 8 = 100 = 23 \bmod~77\]
as the first value, and
\[ \alpha_2 = 4 \times 11 \times 2 + 1 \times 7 \times 8 = 144 = 67 \bmod~77 \]

\noindent
as the second value. 

It is assumed that Alice and Bob have chosen in advance to use
$\alpha =23$ for their communications.

Consider the message
Alice wishes to send is $m=12$. As in the previous case, she finds its 
companions
by multiplying it successively by $\alpha =23$, thus obtaining 34, and 45.
The rank order of her message is 1.

She now performs the cubic transformation on her
message 12, sending Bob $c= 34$ together with the side information of 1. 

Bob find its three cube roots by first calculating
$c^{1/3} = c^7 = 34^7 = 34$. The other cube roots are:

\[ 34\times 23 = 12  \bmod  77\]
and
\[ 34 \times 67 = 45  \bmod 77\]

Bob knows that the message was one of the three
12, 34, and 45. The side information tells him that
the specific message out of this set is 12.

Since there are three solutions, the side information will require 2 bits.

\section{$\phi (n)$ divisible by 9}
The method proposed will actually work even for $\phi(n)$ that is divisible by
$9$. Although each number will now have $9$ cube roots, they may be bunched
together 
in separate groups in a variety of ways.

The 9 cube roots of 1 may be obtained by solving

\begin{equation}
\alpha^9 -1 = 0
\end{equation}

\noindent
which may be simplified to:
\begin{equation}
(\alpha -1) (\alpha^2 + \alpha + 1) (\alpha^6 + \alpha^3 + 1) = 0
\end{equation}


Let the cube roots of 1 be put in numerical order and labeled 1 through 9.
Alice would take the message she wishes to send to Bob and find all the 
multiples
of it with the cube roots of 1, and find its relative position in the set.
Then she will send the original message with this position number as side
information.

Bob will obtain the cube root of $c$ and
then list all its companion solutions. Given the position number, he will
then be able to determine its value.

\vspace{0.1in}
\noindent
{\bf Example 3.} Consider $c = m^3  \bmod 91$, where
$\phi (n) = 72$. Since 88 is square
modulo 91, one can use equation (3) to determine the cube roots
of 1 modulo 91, and a simple calculation gives us the values
9 and 81.  Since 9 and 81 are not cubes themselves, we cannot use
equation (8) to find the remaining cube roots. Additional cube roots 
of 1 are obtained by solving the equation $\alpha^3 -1 = 0$ for the
two moduli 7 and 13 (4 and 10 are square modulo 7 and 13, respectively)
and using CRT to combine the results. We end up with the following
9 cube roots of 1: 
1, 9, 16, 22, 29, 53, 74, 79, 81. 

Assume Alice wishes to send the message 24 (which is relatively
prime to 91, and therefore it has 9 cube roots) to Bob. She finds the multiples
of 24 with the cube roots of 1, and obtains the set: 20, 24, 33, 34, 47, 59, 73, 76, and 89. 
The sequence number of 24 is 2, and this number is sent as side 
information, together with $24^3=83$.

Bob first finds the cube root of 83 by the use of CRT or some probabilistic
algorithm. Suppose, this number is 33. Now he computes the multiples of 33 with
all the cube roots of 1, obtaining: 33, 24, 73, 89, 47, 20, 76, 59, 34. Since 
the side information has revealed that the message has the second rank in this set,
he now knows that it is 24.

\subsection*{Probability events}
The 9 cube roots of 1
may be written as numbers and their squares:

\begin{equation}
1, a, b, c, d, a^2, b^2, c^2, d^2 
\end{equation}

\vspace{0.1in}
These may be bunched together in $3$ groups. Alternatively, 
if it is agreed by Alice and Bob
that each group of three have 1 in it, these numbers may be
put in four groups in the following manner:

\begin{equation}
1, a, a^2; ~~~~ 1, b, b^2; ~~~~ 1, c, c^2; ~~~~ 1, d, d^2 
\end{equation}

\vspace{0.1in}
Thus, in our example, the four sets will be:

\[1, 9, 81;~~~ 1, 16, 74; ~~~ 1, 22, 29; ~~~ 1, 53, 79\]

\vspace{0.1in}
Let the protocol require that Bob pick one of these four sets. Bob
is also told the rank order within each subgroup. His
chances of obtaining
the correct message are $\frac{1}{4}$.

It is clear that the groupings could be done differently as well.
But whichever way the groupings are done, they are 
incorporated in the protocol of the communicating parties.

\section{Generalizations}

Generalizations of this method to higher exponents may be readily made. 

The method of tags may also be applied to the squaring transformation.
Once a square roots $\alpha$
 of 1 (other than 1 or $n-1$) is known, 
the 
full set of solutions follows:

\[1, \alpha, n- 1, n - \alpha\]

One needs 2 bits of side information to uniquely identify the message $m$
that gave rise to the received $c$.

For $k$th-roots of 1 related to the solution to 
$c = m^k \bmod n$, $k$ odd,
the number of groups to generate probability 
events in the manner of equation (10) will be $k+1$, that is
$\frac{k^2 -1}{k-1}$, corresponding to the probability of $\frac{1}{k+1}$.

\subsection*{Random number generation}
The cubic transformation may also be used to generate random numbers
$s_i$ in a manner
analogous to the squaring
transformation [4]. Let $s_0$ be the seed and let $n = pq$ for primes $p$, $q$,
where $\phi (n)$ is divisible by 9,
and $s_0$ is relatively prime to $n$, then,

\begin{equation}
s_i = (s_{i-1})^3 \bmod n
\end{equation}

\vspace{0.1in}
\noindent
The condition of divisibility of $\phi (n)$
by 9 is to increase the computational
burden of inverting the transformation.
The tree of possibilities has 9 branches at each step.

For a cryptographically strong random number generator to radix-$r$ that
generates the sequence $c_i$, one may
reduce the numbers $s_i$ to the modulus $r$, where $r < n$:
\begin{equation}
c_i = s_i \bmod r
\end{equation}

\vspace{0.1in}
\noindent
For binary random numbers generated by this equation, $r$ will be 2.

\section*{References}
\begin{description}

\item
[1]
M.O. Rabin, Digitalized signatures and public-key functions as intractable
as factorization. Technical Report MIT/CCS/TR-212. January 1979.

\item
[2]
R.L. Rivest, A. Shamir, and L. Adleman, A method for obtaining digital
signatures and public-key cryptosystems. Comm. ACM 21: 128-138, 1978.

\item
[3]
P. Garrett, D. Lieman (eds.), Public-Key Cryptography. American Mathematical
Society, Providence, 2005.

\item
[4]
L. Blum, M. Blum, and M. Shub, A simple unpredictable random number
generator. SIAM Journal on Computing 15: 364-383, 1986.

\end{description}
 
\end{document}